\documentclass[twocolumn,superscriptaddress,floatfix,showpacs,aps,prl]{revtex4-1}
\usepackage{mathtext}
\usepackage{graphics}
\usepackage[dvips]{epsfig}
\usepackage{amsmath}

\def\XXint#1#2#3{{\setbox0=\hbox{$#1{#2#3}{\int}$}
\vcenter{\hbox{$#2#3$}}\kern-.55\wd0}}

\newcounter{fig}

\begin{document}
\title{Structure and electron bands of phosphorus allotropes}
\author{L.A. Falkovsky}
\affiliation{ Landau Institute for Theoretical Physics, Chernogolovka 142432}
\affiliation{Verechagin Institute of the High Pressure
Physics, Troitsk 142190}

 \begin{abstract}
The small difference between the rhombohedral phosphorus lattice ($A$-7 phase)  and the simple cubic phase as well as  between phosphorene and the cubic structure is used in order to construct their quasiparticle band dispersion. We  exploit  the Peierls idea
of the Brillouin zone  doubling, which has been previously employed in  consideration of semimetals of the $V$ period and $IV$--$VI$ semiconductors.
 In the common framework,  individual properties of phosphorus allotropes are  revealed.

 \pacs{73.20.AT,73.22.-f,74.20.Pq}
\end{abstract}

\maketitle

{\it Introduction---}

In the last decade much progress in study of the graphene monolayer has been achieved \cite{CN}. Graphene turned out to be a material with remarkable properties: the universal optical conductivity $e^2/4\hbar$, light transmitance giving the fine--structure constant, the Coulomb renormalization of the Fermi velocity. However, absence of a band gap does not permit to use graphene in field effect transistor devices. This is the reason to seek other two dimensional materials with sizable gap and high mobilities. One of such promised substances  becomes phosphorene, i.\,e. a monolayer of black phosphorus.

The element phosphorus belongs to the same period of the periodic table as semimetals As, Sb, and Bi and contains  two  $3s$ and three $3p$ valent electrons. There exist  at least three  phosphorus allotropes: simple cubic, rhombohedral (bismuth $A$-7 symmetry), and orthorhombic ($A$-17). At 4.5 GPa, the structure changes from $A$-17 to $A$-7, which transforms to the sc one at 10 GPa. There are  two phase transformations  also at 137 and 262 GPa. The cubic allotrope possesses one remarkable property - it becomes a superconductor with maximum $T_c$= 9.5 K under pressure 32 GPa \cite{KST,KIE}, and  the high  temperature of the superconducting transition was explained in the Ref. \cite{CMC} by  the electron-phonon  coupling. 

The orthorhombic black phosphorus (BP) is the most stable allotrope of the element. It is a layered material with each layer forming a puckered surface.   Anisotropic optical response \cite{LRW} with a high mobility of carriers promises BP as viable linear polarizers for applications.
The high mobility, tunable bandgap, and linear dichroism along two in-plane directions make few-layer phosphorus a candidate for future electronics and optoelectronics.   

Recently, the success was achieved in fabricating field-effect transistors \cite{LNZ,LYY, BGB,XWJ} based on few-layer BP. Comparing to graphene, BP has an advantage  possessing a quasiparticle band gap.  
The band gap value of monolayer (0.9--1.7 eV in various calculations) tends to smaller values (0.1--0.36 eV) in bulk BP \cite{QKH,GZJ}. The calculations within the local-orbital method \cite{TM,AM} or modern first-principal ones \cite{GZJ,RK,CVP} give the different values of the minimum band gap. That is the direct gap at the $\Gamma$ point or the indirect band  \cite{GLB} between the valence-band maximum at the Z point and the conduction-band  minimum at $\Gamma$.
Sometimes, the variations in the band gap value are explained   by the many-particle correlation effects. According to Ref. \cite{TSL}, the self-energy correction enlarges the band gap from 0.8 to 2 eV, however, the optical absorption peak is reduced to 1.2 eV, that can  be broadly tuned by changing the number of stacked layers.

The spin-orbit interaction $\Delta_{so}$ is pointed out as a possible source of contradictions in the calculations of the band structure. Because the spin-orbit interaction is proportional to the ion charge  squared and
$\Delta_{so}$=1.2 eV  in bismuth, we can estimate  the spin-orbit value for BP as $\Delta_{so}\approx$ 0.04 eV, i.\,e. much smaller in comparison with the interesting energy of the order of 0.2 eV in BP. Therefore, the  spin-orbit coupling cannot noticeably change the quasiparticle band structure in phosphorus \cite{QKH}. 
\begin{figure}[b]
\resizebox{.4\textwidth}{!}
{\includegraphics{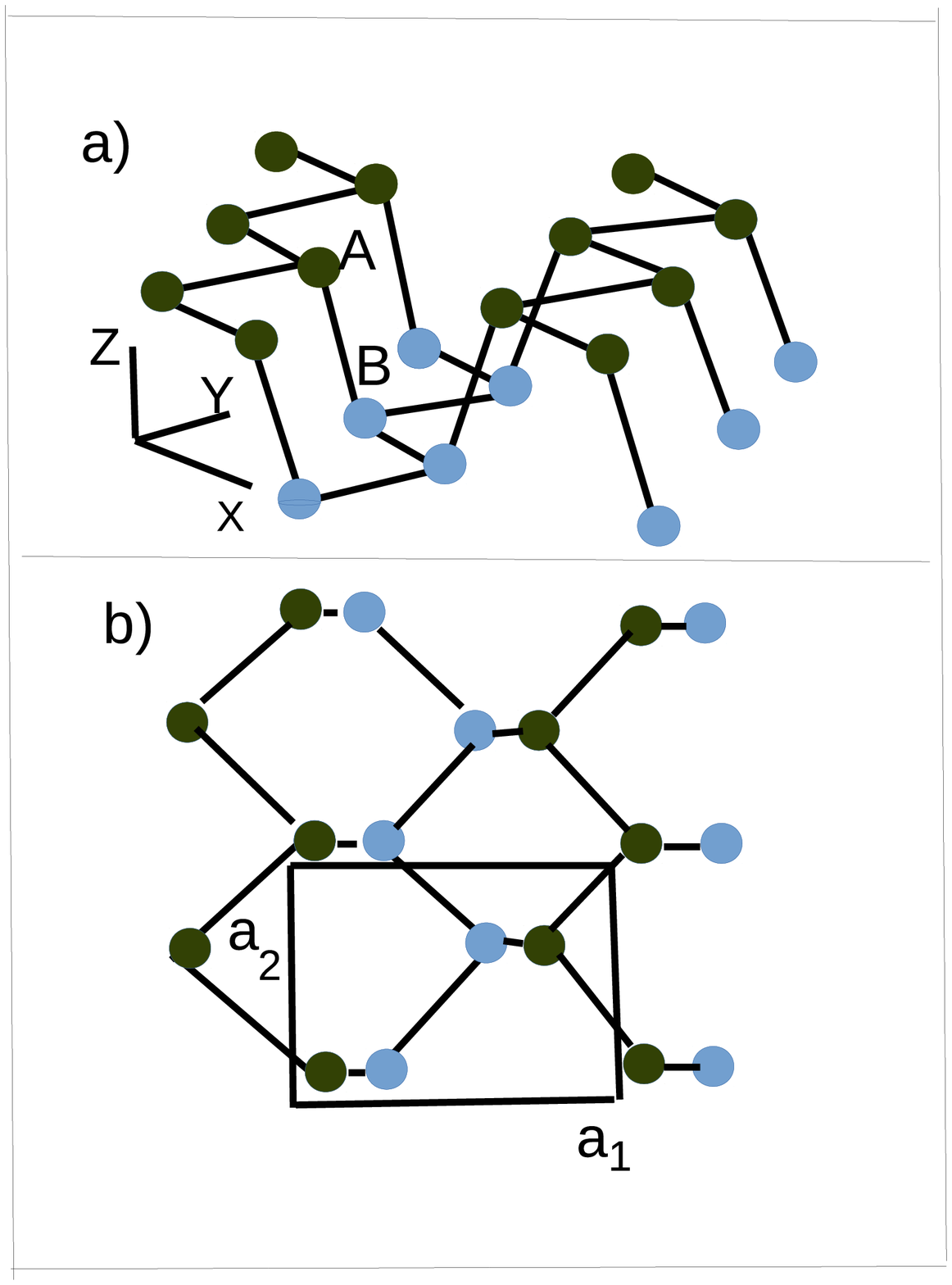}}
\caption{(Color online) Structure of phosphorene. a) 3D representation; the A and B atoms belong to two planes constituting a puckered monolayer phosphoren (see text). b) Top view; the $a_1, a_2$ box shows the unit cell with four atoms.  }
\label{phos}
\end{figure}

The structure of BP  consists of two puckered surfaces. One such the surface is shown in Fig. \ref{phos}.  Each atom is connected with the nearest-neighbors by three bonds. There are \cite{CVP} two bonds of 2.16 $\AA$ and one bond of 2.21 $\AA$, two bond angles are  103.7$^{\circ}$  and one bond angle  is 98.1$^{\circ}$.  There are four atoms in the unit cell, and we have to obtain twelve bands for the $p_i$ orbitals, $i=x, y, z$, hybridized weakly with the $s$ bonds. 

If we set all the bonds be equal and all the angles be equal to 90$^{\circ}$, we obtain two planes of atoms with the simple cubic arrangement instead of one  puckered phosphorene layer. Then we have to take the small  difference between the cubic arrangement and phosphorene  into account. The corresponding deformation is known as  a Peierls  distortion and was employed for the evaluation of the electron dispersion in bismuth as well as in  $IV$-$VI$ compounds \cite{AF,FG,VP,VF}. The similar procedure used for construction of the phonon spectrum is known as folding. 

In this work, we have taken the Peierls idea as a starting point in constructing  an effective low-energy Hamiltonian for  phosphorus allotropes. Within the common framework, the electron dispersion is evaluated in an explicit form for the sc, $A$-7 allotropes, and phosphorene. 

{\it Electron band--dispersion  for a semimetal with the rhombohedral lattice and two atoms in the unit cell---}

Here we consider the electron dispersion of the sc  and $A$-7 lattices for the elements of the $V$ group, starting from the tight-binding approximation and using the Peierls idea of the unit-cell doubling.  There are one atom in the unit cell of the sc lattice and two atoms in the $A$-7 unit cell.
The correct Bloch functions of the zeroth approximation are constructed from the Wannier  orbitals $p_i({\bf r})$  as a sum over the lattice sites $n$ and   atoms $A$ in the unit cell as
\begin{equation}
\psi_{i}({\bf r})=\sum_{n,A}e^{{i\bf kr}_n^A}p_i({\bf r-r}_n^A)\,,
\label{bfu}\end{equation}
where $i=x,y,x$ numerates the  $p_i$ orbitals. We assume that the $s$ orbitals are located more deeply and are separated well from the $p_i$ bands.

\begin{figure}[b]
\resizebox{.4\textwidth}{!}
{\includegraphics{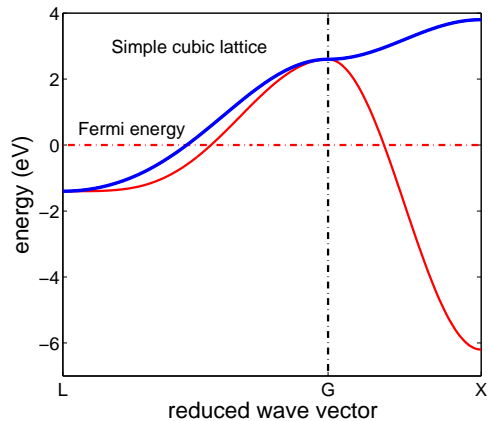}}
\caption{(Color online) Three electron bands for a metal with a simple cubic lattice; the degenerate band is shown in a thick line. }
\label{cu}
\end{figure}

In the sc lattice, an atom with the coordinates $(000)$ has six nearest neighbors at the distance $a$ in $\pm(100),\, \pm(010),$ and \,$\pm(001)$ sites and twelve next-nearest neighbors in the sites like $(110)$. The electron--lattice interaction $H_0$ of cubic symmetry  gives
the following matrix elements between the first neighbors 
\begin{equation}
\xi_{xx}=\xi_0\cos{k_xa}+\xi_1(\cos{k_ya}+\cos{k_za})
\label{ks}\end{equation}
and between the next-nearest neighbors 
\begin{equation}\begin{array}{c}
\eta_{xy}=\eta_0\sin{k_xa}\sin{k_ya}\,,\\ 
\eta_{xx}=\eta_1\cos{k_ya}\cos{k_za}+\eta_2\cos{k_xa}(\cos{k_ya}+\cos{k_za})\,.
\label{et}\end{array}\end{equation}
Here the hopping integrals are
\begin{equation}
\begin{array}{c}
\xi_0=2\langle p_x(000)|H_0|p_x(100)\rangle\,,\\  
\xi_1=2\langle p_x(000)|H_0|p_x(010) \rangle\,,\\ 
\eta_{0}=-4\langle p_x(000)|H_0|p_y(011) \rangle\,,\\ 
\eta_{1}=4\langle p_x(000)|H_0|p_x(011) \rangle\,,\\ 
\eta_{2}=4\langle p_x(000)|H_0|p_x(110) \rangle\,,
\end{array}
\label{xe}\end{equation}
and the arguments of the $p_i$ orbitals indicate the atom coordinates  where these orbitals are centered.
The remaining matrix elements are given by cyclic permutation of the indices.
In Fig. \ref{cu}, the electron dispersion is shown for the sc lattice with $\xi_0=4$ eV, $\xi_1=-1$ eV, $\eta_0=-0.2$ eV, 
$\eta_1=\eta_2=0.2$ eV. If the interatomic distance $a$ is   short enough and the nearest hopping integrals $\xi_0$ and $\xi_1$ have a typical atomic values, such the crystal should be a good metal. Then three valent electrons can occupied only half places in these three bands degenerate twice in the spin. It is not surprising that  phosphorus becomes a superconductor under pressure, when it displays the sc structure.  

\begin{figure}[b]
\resizebox{.4\textwidth}{!}
{\includegraphics{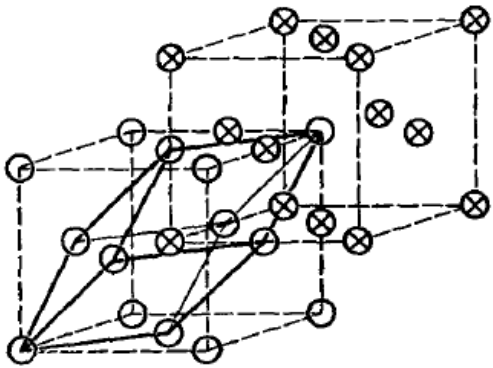}}
\caption{(Color online) Structure of the sc and $A$-7 lattices; two sublattices are shown by  crosses and empty circles, correspondingly. }
\label{a7}
\end{figure}
At high pressure from 40 to 80 kbar, phosphorus  has the rhombohedral structure ($A$-7 phase). 
This structure can be formed from the sc phase by a  small relative displacement of two  face-centered sublattices. Thus, in the unit cell, we get two atoms shown in Fig. \ref{a7} by the cross and empty circles and obtained from the atoms at the  $(000)$ and $(111)$ sites in the cubic phase. In the $A$-7 phase, their rhombohedric  coordinates become $\pm(0.25-u,0.25-u, 0.25-u)$  with $u$ equal to several percents.
For bismuth and rhombohedral phosphorus, such a distortion is on the order  5--10\%. For instance, in bismuth, the angles between bonds become $57^{\circ}$ instead of 60$^{\circ}$ in the sc phase. The Peierls distortion results in the doubling  of the unit cell  volume.
The primitive vectors for the face-centered lattice are 
\[ {\bf a}_i= a(011),\quad a(101), \quad a(110)\,,\]
and the primitive vectors for the reciprocal lattice write
\[ {\bf Q}_i= \pi(-1,1,1)/a,\quad \pi(1,-1,1)/a,\quad \pi(1,1,-1)/a\,.\]

The Peierls distortion can be represented as a result of the  electron--lattice interactions of the rhombohedral symmetry $D_{3d}$, first, due to   the sublattice shift  
\begin{equation}U({\bf r})={\bf u}\cdot \nabla(V_A({\bf r})-V_B({\bf r}))\equiv{\bf u}\cdot {\bf O}\label{u}
\end{equation}
 in the $(111)$ direction and second, because of the deformation 
 \begin{equation}\textit{E}({\bf r})=\varepsilon_{ij}O_{ij}({\bf r})\,.\label{e}
\end{equation} 
 affecting the angles between the bonds and  
 described by the tensor $\varepsilon_{ij}$.

The Hamiltonian for the face-centered lattice with two atoms in the unit cell obtains the form
\begin{equation}
H(\mathbf{k})=\left(
\begin{array}{cc}
A     \, & iU\\
-iU   \, &   A_Q
\end{array}%
\right) ,  \label{ham}
\end{equation}%
where $A$ is the $3\times 3$ matrix given by the matrix elements in Eqs. (\ref{ks})-(\ref{xe}). The matrix $A_Q$ is obtained from $A$ by
the substitution ${\bf k}\rightarrow {\bf k}+{\bf Q}_1+{\bf Q}_2+{\bf Q}_3$. The additional contributions to the matrix $A$  appears from the interaction $E({\bf r})$, Eq. (\ref{e}), and has the matrix elements
 \begin{eqnarray}
E_{xy}=e_0
+e_1(\cos{k_x a}+\cos{k_y a})
+e_2\cos{k_z a}\,,
\label{em}
\nonumber\end{eqnarray}
with the hopping integrals
\begin{equation}
 \begin{array}{c}
e_0=\varepsilon_{xy}\langle p_x(000)|O_{xy}|p_y(000) \rangle\,,\\
e_1=\varepsilon_{xy}\langle p_x(000)|O_{xy}|p_y(100) \rangle\,,\\
e_2=\varepsilon_{xy}\langle p_x(000)|O_{xy}|p_y(001) \rangle\,\,.
\end{array}%
\label{ei}
\end{equation}
The $3\times 3$ matrix $U$ has the matrix elements of the doubling interaction, Eq. (\ref{u}), as following
\begin{equation}
 \begin{array}{c}
U_{xx}=u_1\sin{k_x a}+u_2(\sin{k_y a}+\sin{k_z a})\,,\\
U_{xy}=u_3(\sin{k_x a}+\sin{k_y a})\,,
\end{array}%
\label{um}
\end{equation}
where
\begin{equation}
 \begin{array}{c}
u_1=2\langle p_x(000)|O_x|p_x(100) \rangle\,,\\
u_2=2\langle p_x(000)|O_y|p_y(010) \rangle\,,\\
u_3=2\langle p_x(000)|O_y|p_x(100) \rangle\,.
\end{array}%
\label{ui}
\end{equation}

\begin{figure}[]
\resizebox{.4\textwidth}{!}{\includegraphics{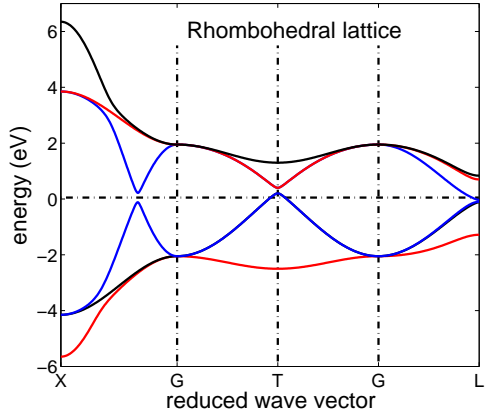}}
\caption{(Color online) Electron dispersion of a rhombohedral lattice with two atoms in unit cell; the holes/electrons are at the $T/L$ points.}
\label{rhom}
\end{figure} 

The Hamiltonian of Eq. (\ref{ham}) gives six bands shown in Fig. \ref{rhom}.
In this case, we get a semimetal  with  holes at the $T$ point and electrons at the $L$ point as it should be in bismuth. The following values of the hopping integral are taken (in eV): $\xi_0=4, \xi_1=-1, u_1=0.2, u_2=0.25, u_3=0.3, \eta_0=-0.3, \eta_1=0.15, \eta_2=-0.1$, and all the parameters of Eq. (\ref{em}) are set to   zero.  Another set of parameters can produce a semiconductor with the narrow band-gap. The largest  hopping integrals $\xi_0$ and $\xi_1$ have approximatively   the similar  values for semimetals of the $V$ period and for the $IV-VI$ semiconductors \cite{VP,VF}. The values of  $e_0, e_1, e_2, u_1, u_2,$ and $u_3$ parameters must do not exceed  0.1$\xi_0$, because they are proportional to the Peierls weak distortion. The spin-orbit interaction should be included for antimony and bismuth, where it is of the same order as the Peierls distortion. 

{\it Electron band--dispersion for phosphorene---}

Now we apply the Peierls doubling to phosphorene. In the simple case of doubling, we obtain the unit cell with two atoms instead of one atom. However, we have now four atoms in the unit cell of phosphorene. Let  three $p_i$ bonds in phosphorene be equal to one  another and the angles between the bonds be equal to 90$^{\circ}$. Thus, we obtain two parallel planes with one atom $A$ under another atom $B$ in these two planes at the distance $a$. The distance between the nearest atoms in the plane is also equal to $a$.   Instead of one monolayer phosphorene, we have two parallel  planes   with two atoms $A$ and $B$  in the unit cell and with the translation symmetry only in $x$ and $y$ directions  (see, Fig. \ref{phos}). Therefore, the vector ${\bf k}$ has only two component $x$ and $y$. Instead of one matrix in the rhombohedral case,  Eqs. (\ref{ks})-(\ref{xe}), we have to combine the matrix $\mathcal{A}$, connecting the nearest and next-nearest neighbors  of the $A$ type, the matrix $\mathcal{B}=\mathcal{A}$, 
connecting the  neighbors  of the $B$ type, and the matrix $\mathcal{A}_B$, connecting the atoms of the $A$ and $B$ type. We obtain the low-energy Hamiltonian  in the form of the $6\times 6$ matrix 
\begin{equation}
A(\mathbf{k})=\left(
\begin{array}{cc}
\mathcal{A}     \, & \mathcal{A}_B\\
\mathcal{A}_B^{\dag} \, &   \mathcal{A}
\end{array}\right)\,.\label{AA}\end{equation}
 The matrix elements of these matrices are obtained as the hopping integrals of the cubic Hamiltonian $H_0$ between the  $p_i$ orbitals centered on two atoms $A$ and $B$: 
\begin{equation}\begin{array}{c}
 \mathcal{A}_{xx}=\xi_0\cos{k_x a}+\xi_1\cos{k_y a}
    +\eta_2\cos{k_x a}\cos{k_y a}\,,\\
 \mathcal{A}_{yy}=\xi_0\cos{k_y a}+\xi_1\cos{k_x a}
    +\eta_2\cos{k_y a}\cos{k_x a}\,,\\
 \mathcal{A}_{zz}=\xi_1(\cos{k_x a}+\cos{k_y a})
    +\eta_1\cos{k_x a}\cos{k_y a}\,,\\
 \mathcal{A}_{xy}= \mathcal{A}_{yx}=\eta_0\sin{k_x a}\sin{k_y a}\,,\\
 \mathcal{A}_{xz}= \mathcal{A}_{zx}= \mathcal{A}_{yz}= \mathcal{A}_{zy}=0\,,\\
 \mathcal{A}_{Bxx}=0.5\eta_1\cos{k_y a}+\eta_2\cos{k_x a}(\cos{k_y a}+0.5)\,,\\
 \mathcal{A}_{Byy}=0.5\eta_1\cos{k_x a}+\eta_2\cos{k_y a}(\cos{k_x a}+0.5)\,,\\
 \mathcal{A}_{Bzz}=0.5\xi_0+\eta_1\cos{k_x a}\cos{k_y a}\\+0.5\eta_2(\cos{k_x a}+\cos{k_y a})\,,\\
 \mathcal{A}_{Bxz}=0.5i\eta_0\sin{k_x a}\,,
 \mathcal{A}_{Byz}=0.5i\eta_0\sin{k_y a}\,,\\
 \mathcal{A}_{Bzx}=- \mathcal{A}_{Bxz}\,,
 \mathcal{A}_{Bzy}=- \mathcal{A}_{Byz}\,,
 \mathcal{A}_{Bxy}= \mathcal{A}_{Byx}=0\,
\end{array}\label{AB}\end{equation}
with  $\xi $ and $\eta$ having the same meaning  as in Eq. (\ref{xe}).
 
 If two primitive vectors of the cubic lattice in the $x,y$ plane are denoted as $a(10)$ and $a(01)$, then the lattice with the doubling unit cell has the primitive vectors ${\bf a}_1=a(1,1)$ and ${\bf a}_2=a(-1,1)$. The primitive vectors of the reciprocal lattice are
\begin{equation} {\bf Q}_1= \pi(1,1)/a,\quad {\bf Q}_2=\pi(-1,1)/a\,.
\label{qu}\end{equation}
Because the vectors ${\bf k}$ and ${\bf k +Q}_1$ are  equivalent in the Brillouin zone after  doubling in the direction ${\bf Q}_1 $ (this is the $x$ direction in phosphorene, see Fig. \ref{phos}), we  have to join  the matrices $A(\mathbf{k})$ and $A(\mathbf{k+Q}_1)$ into the $12\times 12$ matrix. Thus, we get the low-energy effective Hamiltonian for four atoms with the $p_i$ orbitals in the unit cell in the form  of Eq. (\ref{ham}), where the matrix $A_Q$ is obtained
from the matrix of Eqs. (\ref{AA}) and (\ref{AB}) with  the replacement $(\cos{k_x a}, \cos{k_x a})\rightarrow (-\cos{k_x a}, -\cos{k_x a})$. The  $U$ matrix  has the form
\begin{equation}
U(\mathbf{k})=\left(
\begin{array}{cc}
 \mathcal{UA}     \, &  0\\
 0 \, &    \mathcal{UA}
\end{array}\right)\,\label{UU}\end{equation}
with the matrix elements of the orthorhombic Hamiltonian in Eq. (\ref{u}):  
\begin{equation}\begin{array}{c}
 \mathcal{UA}_{xx}=u_1\sin{k_x a}+u_2\sin{k_y a}\,,\\
 \mathcal{UA}_{yy}=u_1\sin{k_y a}+u_2\sin{k_x a}\,,\\
 \mathcal{UA}_{zz}=u_2(\sin{k_y a}+\sin{k_x a})\,,\\
 \mathcal{UA}_{yx}= \mathcal{UA}_{xy}=u_3(\sin{k_x a}+\sin{k_y a})\,,\\
 \mathcal{UA}_{zx}= \mathcal{UA}_{xz}=u_3\sin{k_x a}\,,\\
 \mathcal{UA}_{zy}= \mathcal{UA}_{yz}=u_3\sin{k_y a}\,
\end{array}\label{UAB}\end{equation}

The Hamiltonian of Eq.  (\ref{e}) has  the orthorhombic symmetry $C_{2h}$  as well as the  Hamiltonian in Eq. (\ref{u}). The corresponding   matrix elements 
 \begin{equation}
 \begin{array}{c}
E_{xy}=e_0+e_1(\cos{k_x a}+\cos{k_y a})\,,\\%
E_{yz}=e_0+e_1\cos{k_y a}+e_2\cos{k_x a}\,,\\
E_{xz}=e_0+e_1\cos{k_x a}+e_2\cos{k_y a}\,
\end{array}%
\label{e1}
\end{equation}%
 are added to the matrix $\mathcal{A}$. 

\begin{figure}[]
\resizebox{.4\textwidth}{!}{\includegraphics{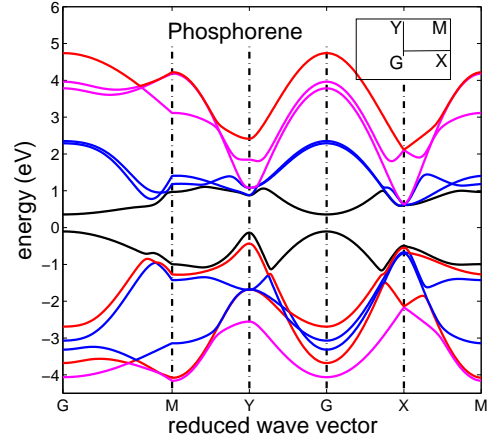}}
\caption{(Color online) Electronic dispersion for the black phosphorus monolayer.}
\label{bp}
\end{figure}

The band structure  for phosphorene is shown in Fig. \ref{bp} with the following values of the hopping integrals (in eV):
$\xi_0=4.1, \xi_1=-1.0,\eta_0=0.1, \eta_1=0.3, \eta_2=-0.1, u_1=0.4, u_2=0.09, u_3=-0.1,  e_0=0.2, e_1=0.25, e_2=0.15$.
This set gives a semiconductor with the minimal quasiparticle band gap  at the $\Gamma$ point  $\varepsilon_g$= 0.46 eV. All the energies at $\Gamma$ are expressed in terms of the above hopping integrals (see Supplemental Material at \cite{bp_Sl}). The band dispersion  is more flat along the $\Gamma-M$ direction than in the $\Gamma -X$ and $\Gamma -Y$ directions.
The effective masses of holes do not coincide with the masses of electrons \cite{RCC}, because their dispersions are described by  rather  independent equations.
  At the $\Gamma$ point,  the conduction band  is  determined  by the matrix $A({\bf k})$ in Eqs. (\ref{AA}), (\ref{AB}) and  the valence band corresponds to the matrix $A({\bf k+Q}_1)$. The matrix elements between them are the terms with $u_i$, Eqs. (\ref{UU}) and (\ref{UAB}). The last elements  vanish at the $\Gamma$ point and should be expanded in  $k_x, k_y$ in order to find the effective masses. Using the foregoing set of parameters, we obtain for the inverse effective masses $m^{-1}$ the sum of two terms   $ a^2(\xi_0+\xi_1)/\hbar^2\pm a^2u_1^2/\hbar^2\varepsilon_g$ for  electrons and holes, correspondingly. Here, the first term plays a main role. Thus, we estimate  the masses as $m\simeq  0.5 m_0$.

{\it Summary---}

At 4.5 GPa, the phosphore structure changes from the orthorhombic  ($A$-17)   to rhombohedral ($A$-7, of bismuth type), which transforms to the sc one at 10 GPa. The structure of the orthorhombic and rhombohedral phases differs slightly from the more symmetrical sc structure. Therefore, their  quasiparticle dispersion  can be obtained in the framework of  the general Peierls idea of the doubling distortion. We show that in  agreement with  experiments, the  low energy Hamiltonian constructed in accordance with the Peierls method gives the  dispersion of a metal for the sc phase,  of a semimetal or a narrow gap semiconductor for the $A$-7 phase, and of a semiconductor for phosphorene. Because of the small gap at different points in the Brillouin zone ($\Gamma$, $X$, and $Y$), phosphorene can be transformed from  direct band gap semiconductor to indirect semiconductor or semimetal with the compression or in the multilayer structures.

This research was initiated by V. Brazhkin who called the author's attention to the problem of phosphorene. 
We acknowledge  the Russian Foundation for Basic
Research (grant No. 13-02-00244A) and the SIMTECH Program, New Centure of Superconductivity: Ideas, Materials and Technologies (grant No. 246937).


\pagebreak
\begin{center}
\textbf{\large Supplemental Material for\\
``Structure and electron bands of phosphorus allotropes''}
\end{center}
\setcounter{equation}{0}
\setcounter{page}{1}
\makeatletter
\renewcommand{\theequation}{S\arabic{equation}}

\section{Band energies in phosphorene at the $\Gamma$ point}

At the $\Gamma $ point, the matrix $U({\bf k})$, Eqs. (14-15), vanishes and the band energies are determined by the eigenvalues of two matrices, $A({\bf k}=0)$, Eqs. (11-12), and  $A({\bf Q}_1)$. In addition,
the matrix $\mathcal{A}_B$ in  Eq. (12) takes the diagonal form with the elements $\mathcal{A}_{Bxx}=\mathcal{A}_{Byy}=0.5\eta_1+1.5\eta_2\,,
 \mathcal{A}_{Bzz}=0.5\xi_0+\eta_1+\eta_2 $. Therefore, we can find three elements of the eigenfunction $(\varphi_4,\varphi_5,\varphi_6)$ from the first three eigen-equations, given by the matrix of Eq. (11), and substitute them in the second three eigen-equations. As a result, we get the equation determining the energies at the $\Gamma$ point in the form
\begin{equation}
\left|
\begin{array}{ccc}
d_{11}     \, & d_{12}\, & d_{13}\\
d_{12} \, & d_{11}\, & d_{13}\\
d_{13}\, & d_{13}\, & d_{33}
\end{array}\right|=0\,,\label{d}\end{equation}
where
\begin{equation}\begin{array}{c}
d_{11}=\displaystyle{\mathcal{A}_{Bxx}-\frac{(\mathcal{A}_{xx}-\varepsilon)^2+\mathcal{A}_{xy}^2}{\mathcal{A}_{Bxx}}-\frac{\mathcal{A}_{xz}^2}{\mathcal{A}_{Bzz}}}\,,\\
d_{12}=-2\displaystyle{\mathcal{A}_{xy}\frac{\mathcal{A}_{xx}-\varepsilon}{\mathcal{A}_{Bxx}}-\frac{\mathcal{A}_{xz}^2}{\mathcal{A}_{Bzz}}}\,,\\
d_{13}=-\displaystyle{\mathcal{A}_{xz}\left[\frac{\mathcal{A}_{xx}+\mathcal{A}_{xy}-\varepsilon}{\mathcal{A}_{Bxx}} +\frac{\mathcal{A}_{xx}-\varepsilon}{\mathcal{A}_{Bzz}} \right]}\,,\\
d_{33}=\displaystyle{\mathcal{A}_{Bzz}-\frac{(\mathcal{A}_{zz}-\varepsilon)^2}{\mathcal{A}_{Bzz}} -2\frac{\mathcal{A}_{xz}^2}{\mathcal{A}_{Bxx}}}\,.
\end{array}\label{dd}\end{equation}

The equation (\ref{d}) is reduced to two equations
\begin{equation}
d_{11}-d_{12}=0\quad \text{and} \quad (d_{11}+d_{12})d_{33}-2d_{13}^2=0.
\label{d2}\end{equation}
The first quadratic equation gives two band energies
\begin{equation}\varepsilon_{1,2}=\mathcal{A}_{xx}-\mathcal{A}_{xy}\pm \mathcal{A}_{Bxx}\label{eq1}\end{equation}
The second equation of the fourth order also can be solved giving the energies
\begin{equation}
\begin{array}{c}\varepsilon_{3,4}=0.5(\mathcal{A}_{xx}+\mathcal{A}_{xy}+\mathcal{A}_{zz}+\mathcal{A}_{Bxx}+\mathcal{A}_{Bzz})\\ \pm[ 0.25(\mathcal{A}_{xx}+\mathcal{A}_{xy}-\mathcal{A}_{zz}+\mathcal{A}_{Bxx}-\mathcal{A}_{Bzz})^2+2\mathcal{A}_{xz}^2]^{1/2}\end{array}\label{eq2}\end{equation}
and $\varepsilon_{5,6}$  obtained with changing the sigh at $\mathcal{A}_{Bxx}$ and $\mathcal{A}_{Bzz}$. The same set of parameters as in Fig. 5 give for the  $\varepsilon_{1}\div\varepsilon_{6}$ energies (in eV):
2.35,  2.29, 3.78, -4.06,  3.96,  0.36.  

Other six bands at the $\Gamma$ point have the opposite parity. Their energies are given in Eqs. (\ref{eq1}) and (\ref{eq2}) with changing the sign of $\xi_0$ and $\xi_1$.
We obtain (in eV):
-3.07, -2.69, -0.10, -3.3151,  4.74, -3.68. 
\end{document}